%%
%% last modified at Chicago, Sep. 10,
%%

%-------------------------
% This paper uses harvmac
%-------------------------
\input harvmac.tex
\noblackbox
%

% Something to deal with sub-sub-sections

\def\unlockat{\catcode`\@=11}
\def\lockat{\catcode`\@=12}

\unlockat
% Something to deal with sub-sub-sections

\def\newsec#1{\global\advance\secno by1\message{(\the\secno. #1)}
\global\subsecno=0\global\subsubsecno=0\eqnres@t\noindent
{\bf\the\secno. #1}
\writetoca{{\secsym} {#1}}\par\nobreak\medskip\nobreak}
\global\newcount\subsecno \global\subsecno=0
\def\subsec#1{\global\advance\subsecno
by1\message{(\secsym\the\subsecno. #1)}
\ifnum\lastpenalty>9000\else\bigbreak\fi\global\subsubsecno=0
\noindent{\it\secsym\the\subsecno. #1}
\writetoca{\string\quad {\secsym\the\subsecno.} {#1}}
\par\nobreak\medskip\nobreak}
\global\newcount\subsubsecno \global\subsubsecno=0
\def\subsubsec#1{\global\advance\subsubsecno by1
\message{(\secsym\the\subsecno.\the\subsubsecno. #1)}
\ifnum\lastpenalty>9000\else\bigbreak\fi
\noindent\quad{\secsym\the\subsecno.\the\subsubsecno.}{#1}
\writetoca{\string\qquad{\secsym\the\subsecno.\the\subsubsecno.}{#1}}
\par\nobreak\medskip\nobreak}

\def\subsubseclab#1{\DefWarn#1\xdef
#1{\noexpand\hyperref{}{subsubsection}%
{\secsym\the\subsecno.\the\subsubsecno}%
{\secsym\the\subsecno.\the\subsubsecno}}%
\writedef{#1\leftbracket#1}\wrlabeL{#1=#1}}% Macros for boxes
\lockat

%
%-------------------
%  definitions
%-------------------
%
%
\def\CC{{\cal C}}

\def\CG{{\cal G}}
\def\CH{{\cal H}}

\def\CM{{\cal M}}
\def\CN{{\cal N}}

\def\IZ{\relax\ifmmode\mathchoice
{\hbox{\cmss Z\kern-.4em Z}}{\hbox{\cmss Z\kern-.4em Z}}
{\lower.9pt\hbox{\cmsss Z\kern-.4em Z}}
{\lower1.2pt\hbox{\cmsss Z\kern-.4em Z}}\else{\cmss Z\kern-.4em
Z}\fi}
\def\IB{\relax{\rm I\kern-.18em B}}
\def\IC{{\relax\hbox{$\inbar\kern-.3em{\rm C}$}}}
\def\ID{\relax{\rm I\kern-.18em D}}
\def\IE{\relax{\rm I\kern-.18em E}}
\def\IF{\relax{\rm I\kern-.18em F}}
\def\IG{\relax\hbox{$\inbar\kern-.3em{\rm G}$}}
\def\IGa{\relax\hbox{${\rm I}\kern-.18em\Gamma$}}
\def\IH{\relax{\rm I\kern-.18em H}}
\def\II{\relax{\rm I\kern-.18em I}}
\def\IK{\relax{\rm I\kern-.18em K}}
\def\IP{\relax{\rm I\kern-.18em P}}

\def\inbar{\,\vrule height1.5ex width.4pt depth0pt}
\def\p{\partial}

\font\cmss=cmss10 \font\cmsss=cmss10 at 7pt
\def\IR{\relax{\rm I\kern-.18em R}}

\def\Tr{\rm Tr}

\def\CZ{{\cal Z}}

\def\lala{\langle \langle}
\def\rara{\rangle\rangle}
\def\ra{\rangle}
\def\la{\langle}

%%% MACROS FOR BOX BOUNDARY CONDS
%%% FROM KAWAI ET AL

\def\makeblankbox#1#2{\hbox{\lower\dp0\vbox{\hidehrule{#1}{#2}%
   \kern -#1% overlap rules
   \hbox to \wd0{\hidevrule{#1}{#2}%
      \raise\ht0\vbox to #1{}% vrule height
      \lower\dp0\vtop to #1{}% vrule depth
      \hfil\hidevrule{#2}{#1}}%
   \kern-#1\hidehrule{#2}{#1}}}%
}%
\def\hidehrule#1#2{\kern-#1\hrule height#1 depth#2 \kern-#2}%
\def\hidevrule#1#2{\kern-#1{\dimen0=#1\advance\dimen0 by #2\vrule
    width\dimen0}\kern-#2}%
\def\openbox{\ht0=1.2mm \dp0=1.2mm \wd0=2.4mm  \raise 2.75pt
\makeblankbox {.25pt} {.25pt}  }
\def\opensquare{\ht0=3.4mm \dp0=3.4mm \wd0=6.8mm  \raise 2.7pt \makeblankbox
{.25pt} {.25pt}  }

\def\sector#1#2{\ {\scriptstyle #1}\hskip 1mm
\mathop{\opensquare}\limits_{\lower 1mm\hbox{$\scriptstyle#2$}}\hskip 1mm}

\def\tsector#1#2{\ {\scriptstyle #1}\hskip 1mm
\mathop{\opensquare}\limits_{\lower 1mm\hbox{$\scriptstyle#2$}}^\sim\hskip 1mm}
%%%
%%%

%
%-------------------
% references
%-----------------

\lref\sennon{A. Sen,``NonBPS States and Branes in String Theory,''
 hep-th/9904207.}

\lref\fks{S. Ferrara, R. Kallosh and A. Strominger, ``N=2 Extremal
Black Holes,'' Phys. Rev. {\bf D52} (1995) 5412, hep-th/9508072.}

\lref\mli{M. Li, Nucl. Phys. {\bf B460} (1996) 351; hep-th/9510161.}

\lref\gg{M. B. Green and M. Gutperle, Nucl. Phys. {\bf B476} (1996) 484;
hep-th/9604091.}

\lref\dh{L. Dixon and J. A. Harvey, Nucl. Phys. {\bf B274} (1986) 93.}

\lref\sw{N. Seiberg and E. Witten, Nucl. Phys. {\bf B276} (1986) 272.}

\lref\bcr{M. Billo, B. Craps and F. Roose, ``On D-Branes in Type 0 String
Theory,'' Phys. Lett. {\bf B457} (1999) 61; hep-th/9902196.}

\lref\cds{A. Connes, M.R. Douglas, and
A. Schwarz, ``Noncommutative geometry and
Matrix Theory: Compactification on Tori,''
hep-th/9711162.}

\lref\piosch{B. Pioline and A. Schwarz,
``Morita equivalence and $T$-duality
(or $B$ versus $\Theta$),'' hep-th/9908019.}

\lref\ghj{F.M. Goodman, P. de la Harpe,
and V.F.R. Jones, {\it Coxeter Graphs and
Towers of Algebras}, MSRI publications 14,
Springer-Verlag 1989.}

\lref\affludb{I. Affleck and A. W. W. Ludwig, Phys. Rev. {\bf B48} 
(1993) 7297.}

\lref\callkleb{C. G. Callan, Jr. and I. R. Klebanov, ``D-Brane Boundary
State Dynamics,'' Nucl. Phys. {\bf B465} (1996) 473; hep-th/9511173.}

\lref\ers{S. Elitzur, E. Rabinovici and G. Sarkissian,``On Least Action
D-Branes,'' Nucl. Phys. {\bf B541} (1999) 731; hep-th/9807161.}

\lref\affleckludwig{I. Affleck and A.W.W. Ludwig,
``Universal noninteger ground state degeneracy in
critical quantum systems,''
 Phys.Rev.Lett.{\bf 67} (1991) 161.}

\lref\behrend{R.E. Behrend, P.A. Pearce, V.B. Petkova,
J-B Zuber, ``Boundary conditions in Rational
Conformal Field Theories,'' hep-th/9908036.}

\lref\bg{O. Bergman and M. Gaberdiel, ``A Non-supersymmetric
Open String Theory and S-duality'', Nucl. Phys. {\bf B499} (1997)
183, hep-th/9701137.}

\lref\brunner{I. Brunner, M. Douglas,
A. Lawrence, C. Romelsberger, ``D-branes on
the Quintic,'' hep-th/9906200.}

\lref\callan{C. G. Callan, C. Lovelace, C. R. Nappi and S. A. Yost,
``Adding Holes and Corsscaps to the Superstring,'' Nucl. Phys.
{\bf B293} (1987) 83; C.G. Callan, C. Lovelace, C. R. Nappi and S. A. Yost,
``Loop Corrections to Superstring Equations of Motion,'' Nucl. Phys.
{\bf B308} (1988) 221.}

\lref\cardy{J. Cardy, ``Boundary conditions, fusion
rules and the Verlinde formula,'' Nucl. Phys. B324 (1989)581;
``Boundary conditions in conformal field theory,'' Talk at
Taniguchi Symposium on Integrable models, October 1988.}

\lref\cmnpii{A. Cohen, G. Moore, P. Nelson, and
J. Polchinski, ``An invariant string propagator,''
in {\it Unified String Theories}, p. 568,  M. Green
and D. Gross, eds. World Scientific, 1986. }

\lref\dvvv{R. Dijkgraaf et. al.
``The operator algebra of orbifold
models,'' Commun. Math. Phys. {\bf 123}(1989)485.}
\lref\dv{R. Dijkgraaf and E. Verlinde,
``Modular invariance and the fusion algebra,''
Presented at Annecy Conf. on Conformal Field Theory, Annecy, France, Mar 14-16,
1988.
Published in Annecy Field Theory 1988:0087}

\lref\dglmr{M. Douglas and G. Moore}

\lref\ght{G. W. Gibbons, G. T. Horowitz and P. K. Townsend,
``Higher-dimensional resolution of  dilatonic black-hole
singularities,'' Class. Quantum Grav. {\bf 12} (1995) 297.}
\lref\dgt{M. J. Duff, G. W. Gibbons and P. K. Townsend,
``Macroscopic superstrings as interpolating solitons,''
Phys. Lett. {\bf B332} (1994) 321.}

\lref\frsh{D. Friedan and S. Shenker, ``The integrable analytic
geometry of quantum string,'' Phys. Lett. {\bf 175B}(1986) 287;
``The analytic geometry of two dimensional conformal
field theory,'' Nucl. Phys. {\bf B281}(1987) 509;
D. Friedan, ``The space of conformal field theories and the
space of classical string ground states,'' in
{\it Physics and Mathematics of Strings}, L. Brink, D. Friedan,
and A.M. Polyakov eds., World Scientific 1990}

\lref\friedan{D. Friedan, ``The space of conformal
field theories for the c=1 Gaussian model, ''
unpublished, c. 1994. }
\lref\naturality{G. Moore and N. Seiberg,
``Naturality in conformal field theory,'' Nuc. Phys.
{\bf B313} (1989) 16}
\lref\ms{G. Moore and N. Seiberg,
``Classical and Quantum Conformal Field Theory,''
Commun. Math. Phys. {\bf 123}(1989)177;
``Lectures on Rational Conformal Field Theory,''
in {\it Strings '89},Proceedings
of the Trieste Spring School on Superstrings,
3-14 April 1989, M. Green, et. al. Eds. World
Scientific, 1990}
\lref\cardy{J. Cardy, ``Boundary conditions,
fusion rules, and the Verlinde formula,''
Nucl. Phys. {\bf B324} (1989) 581. }
\lref\ishibashi{N. Ishibashi, ``The Boundary and Crosscap
States in Conformal Field Theories'', {\it Mod. Phys. Lett.}
{\bf A4} (1989) 251; N. Ishibashi and T. Onagi,
``Conformal Field Theories on Boundaries and Surfaces
with Crosscaps'', {\it Mod. Phys. Lett.}
{\bf A4} (1989) 161.}
\lref\lewellen{D. Lewellen, ``Sewing constraints
for conformal field theories on surfaces with
boundaries,'' Nucl.Phys. {\bf B372} (1992) 654;
J.L. Cardy and D.C. Lewellen, ``Bulk and boundary
operators in conformal field theory,''
Phys.Lett. {\bf B259} (1991) 274.}
\lref\sagnotti{G. Pradisi, A. Sagnotti and Ya. S. Stanev,
``Planar Duality in $SU(2)$ WZW Models,'' Phys. Lett. {\bf B354}
(1995) 279; hep-th/9503207 \semi G. Pradisi A. Sagnotti and
Ya. S. Stanev, ``The Open Descendants of Nondiagonal $SU(2)$
WZW Models,'' Phys. Lett. {\bf B356} (1995) 230; hep-th/9506014 \semi
G. Pradisi, A. Sagnotii,
and Ya. S. Stanev, ``Completeness conditions
for boundary operators in 2D conformal
field theory,'' Phys. Lett. {\bf B381} (1996) 97; hep-th/9603097.}
\lref\reck{A. Recknagel and V. Schomerus,
``D-branes in Gepner models,'' Nucl. Phys. {\bf B531} (1998) 185;
hep-th/9712186.}
\lref\fs{J. Fuchs and C. Schweigert,
``A classifying algebra for boundary conditions,'' hep-th/9708141,
Phys.Lett. B414 (1997) 251-259 \semi
J. Fuchs and C. Schweigert,
 ``Branes: from free fields to general
backgrounds,'' Nucl. Phys. {\bf B530} (1998) 99, hep-th/9712257\semi
J. Fuchs and C. Schweigert, ``Symmetry breaking boundaries I. General
theory,'' hep-th/9902132 \semi J. Fuchs and C. Schweigert, ``Symmetry
Breaking Boundaries. 2. More Structures:Examples,'' hep-th/9908025.}
\lref\greengut{M. Green and M. Gutperle, ``Light Cone Supersymmetry and
D-Branes,'' Nucl. Phys. {\bf B476} (1996) 484, hep-th/9604091.}
\lref\segalstanford{G. Segal, Stanford lectures}
\lref\divecchia{P. Di Vecchia, M. Frau, I. Pesando,
S. Sciuto, A. Lerda, and R. Russo, ``Classical
p-branes from boundary state,'' hep-th/9707068.}
\lref\jp{J. Polchinski, {\it String Theory},
Cambridge Univ. Press, 1998}
\lref\joed{J. Polchinski, ``Dirichlet Branes and Ramond-Ramond Charges,''
Phys.Rev.Lett. {\bf 75} (1995) 4724.}
\lref\polcai{
J. Polchinski and Y. Cai, ``Consistency of Open Superstring Theories,''
Nucl. Phys. {\bf B296} (1988) 91.}
\lref\nsv{K.S. Narain, M.H. Sarmadi (Rutherford), C. Vafa,
``Asymmetric Orbifolds,'' Nucl.Phys.B288:551,1987 ;
``Asymmetric orbifolds: path integral  and
operator formalism,'' Nucl.Phys.B356:163-207,1991 }
\lref\joenotes{J. Polchinski, ``TASI Lectures on D-branes'',}
\lref\RS{A. Recknagel and V. Schomerus, ``D-branes
in Gepner Models'', hep-th/9712186 \semi
A. Recknagel and V. Schomerus,
``Boundary Deformation Theory and Moduli Spaces of D-Branes,''
hep-th/9811237,Nucl.Phys. B545 (1999) 233-282 \semi
A. Recknagel and V. Schomerus,
``Moduli Spaces of D-branes in CFT-backgrounds,''
hep-th/9903139 .}
\lref\arthatt{G. Moore, ``Arithmetic and attractors,'' hep-th/9807087;
``Attractors and arithmetic,'' hep-th/9807056.}
\lref\fgk{S. Ferrara,  G. W. Gibbons,  R. Kallosh,
``Black Holes and Critical Points in Moduli Space,''  hep-th/9702103}

%
%-------------------
% title page
%-------------------
%
\Title{\vbox{\baselineskip12pt
\hbox{hep-th/9909072}
\hbox{EFI-99-35}
\hbox{YCTP-22-99}
\hbox{SU-ITP-99/42}
\hbox{SLAC-PUB-8245}
}}
{\vbox{\centerline{Tension is Dimension }
 }}

\bigskip
\centerline{Jeffrey A. Harvey$^{\clubsuit}$, Shamit Kachru$^{\diamondsuit}$,
 Gregory Moore$^{\spadesuit,}$\foot{Address after Jan. 1, 2000:
Dept. of Physics and Astronomy, Rutgers University,
Piscataway, NJ 08855-0849} and
Eva  Silverstein$^{\diamondsuit}$}
\bigskip
\centerline{${}^\clubsuit$ Enrico Fermi
Institute and Department of Physics}
\centerline{University of
Chicago, 5640 Ellis Avenue, Chicago, IL 60637}
\bigskip
\centerline{${}^\diamondsuit$ Department of Physics and SLAC}
\centerline{Stanford University}
\centerline{Stanford, CA 94305/94309}
\bigskip
\centerline{${}^\spadesuit$ Department of Physics, Yale University,}
\centerline{Box 208120, New Haven, CT 06520}
\bigskip
\centerline{\bf Abstract}
\noindent
We propose a simple universal formula for the
tension of a D-brane in terms of a regularized
dimension of the associated conformal field
theory statespace.

\Date{September 10, 1999}
%\draft
%
%----------------------
% Body of Paper
%----------------------

\newsec{Introduction}

D-branes are destined to play  a fundamental role in
the formulation of nonperturbative string theory.
Nevertheless, despite much work, and a good
understanding of  examples such as
toroidal compactification, there is
as yet no general formulation of the D-brane
spectrum applicable to an arbitrary closed string
background. In this note we point out one simple
aspect of D-branes which, we conjecture, is
quite general. Namely, the square of the  tension of a D-brane
is proportional  to the regularized dimension of
some infinite dimensional algebra.  We show
that this is indeed true for the wide class of D-branes
associated to rational and quasi-rational conformal
field theories.

While the general construction of the D-brane
spectrum has yet to be carried out, much
is indeed known thanks to recent vigorous
development of boundary conformal field
theory, and the boundary state formalism. A partial list
of references includes
\refs{\callan,\polcai,\cardy,\lewellen,\friedan,
\sagnotti,\mli,\divecchia,\gg,\bg,\fs,\RS,\brunner,\behrend}.
One point which is well-established
is that boundary states $\vert B \rara$
are linear functionals on the closed
string statespace $\CH^{\rm closed}$ preserving
conformal invariance
\eqn\vircond{
(L_n-\tilde L_{-n}) |B\rara =0
}
where $L_n, \tilde L_n$ are the left and right
Virasoro generators.

One might think that \vircond, which imposes conformal
invariance on the string worldsheet, is the only condition one
needs
to impose to find physically acceptable boundary states in
string theory.  Taking this point of view
immediately leads to
problems:
there are far too many solutions to
\vircond\ for an acceptable D-brane spectrum.
Indeed, \vircond\ is a linear equation and solutions
are in one to one correspondence with  spinless
Virasoro primaries $\phi^\alpha_{h,\tilde h}$ in the
closed string
spectrum \cmnpii (here
$h=\tilde h$ and $\alpha$ is a degeneracy index.)

This result is most naturally understood as follows \refs{\friedan,\behrend}.
We can decompose $\CH^{\rm closed}$ in terms of Virasoro  irreps $V_{h}$ as
\eqn\hdecom{\CH^{\rm closed} = \oplus_{(h, \tilde h) \in {\rm Spec}}
 V_h \otimes
\tilde  V_{\tilde h}}
where the spectrum of the closed string theory is determined by the
set Spec of pairs of conformal dimensions. We can then  solve
\vircond\ in each component of $\CH^{\rm closed}$ as
\eqn\bdecom{| B \rangle \rangle_{h,\tilde h} = \sum_{m,\tilde m}a_{m,\tilde m}
 |h,m \rangle \otimes
|\tilde h , \tilde m \rangle \in V_h \otimes \tilde V_{\tilde h} . }
Using the inner product on $\tilde V_{\tilde h}$ under which
$L_n^\dagger = L_{-n}$, $| B \rara_{h,\tilde h}$
 is equivalent to a homomorphism
$B_{h,\tilde h}$ : $\tilde V_{\tilde  h} \rightarrow V_{h}$ given by
\eqn\bdefn{B_{h,\tilde h} = \sum_{m,\tilde m}a_{m,\tilde m}
 |h,m \rangle \otimes \langle \tilde h,\tilde m| }
obeying $L_n B_{h,\tilde h} = B_{h,\tilde h} \tilde L_n$.
That is, $B_{h,\tilde h}$ is an intertwiner between
$V_{h}$ and $\tilde V_{\tilde h}$. Since these are
irreps it follows from Schur's lemma that
  that $|B \rara_{h,\tilde h}$ vanishes if
$h \ne \tilde h$ and that $B_{h,\tilde h}$ is proportional to the identity when
$h=\tilde h$ (the fact that we need
 $h=\tilde h$ is already clear from
the $n=0$ component of \vircond ).
Choosing the proportionality constant
to be one, we can take $B$ acting on $\CH^{\rm closed}$ to be
the projection operator onto $V_h \otimes \tilde V_{\tilde h=h}$.
We denote the corresponding ``states'' in
$\CH^{\rm closed}$ by $|h,\tilde h=h \rangle \rangle$.
There is a natural generalization of this construction to arbitrary
chiral algebras \refs{\ishibashi,\RS,\behrend} and in this context the
states are often referred to as ``Ishibashi
states.'' We will refer to them as ``character
states.''

One problem with these Virasoro character states
is that most of them do not couple
to the graviton at leading order in string
perturbation theory.   This
follows because the overlap with the
graviton state of
a primary or descendent in another irrep (not containing
the graviton state as primary) is zero.    Therefore, if one wrapped
these ``branes" on cycles the resulting particles would have
string scale $\CO(g_s^0)$ masses and one would expect
severe problems with unitarity.

On the other hand, it has been well appreciated for some time
that in addition to \vircond\ various sewing and locality
conditions should be imposed \refs{\lewellen,\sagnotti},
at least if one desires a description in terms of
local boundary conformal field theory on the string
worldsheet.  An especially
important role is played by the Cardy condition \cardy. To state
this condition consider possible boundary states $| \alpha \rara$,
$| \beta \rara $, and compute the partition function (cylinder amplitude)
\eqn\zcyl{Z_{\alpha\beta} = \la\la \beta \vert q_c^{{1\over 2}(L_0^c
+ \tilde L_0^c - {c\over 12})}\vert \alpha \ra\ra}
where $L_0^c, \tilde L_0^c$ are the left and right-moving
closed string Hamiltonians and $q_c = e^{-2 \pi t_c}$.
This can be given a Hamiltonian interpretation
in the open string channel by viewing the cylinder as an annulus with Euclidean
time running around the annulus. After a conformal rescaling of
coordinates  we should thus
be able to write \zcyl\ as
\eqn\zcard{{\rm Tr}_{H_{\alpha \beta}} q_o^{L_0^{\rm open}-c/24}}
where $H_{\alpha \beta}$ is the Hilbert space of open strings with boundary
conditions defined by $\alpha, \beta$ and
 $q_o=e^{-2 \pi t_o} = e^{-2 \pi/t_c}$.
Cardy's condition follows from the equality of \zcyl\ and \zcard. In
other words, the modular transform of \zcyl\ to the variable $q_o$ should
have a $q_o$ expansion with non-negative integer coefficients for all possible
pairs of boundary states $| \alpha \rara, | \beta \rara$. We will call
such a set of boundary states a Cardy set.

In the following we demonstrate that imposing the Cardy condition
reduces the solutions to \vircond\ down to
an acceptable few:  boundary states satisfying the Cardy condition
must couple to the graviton at leading order in string perturbation
theory, and therefore have tensions of order $\CO(g_s^{-1})$.

Moreover, we find that
the tensions are given by the (suitably regularized) dimension of
an associated open string statespace as in equations
 (2.3) and (3.13) below. We
 believe that the generality of this result has not  been appreciated
previously, although many of the elements of our argument are not
new. In particular,
the  regularized dimension
has appeared previously in the literature on boundary CFT as the
boundary entropy of Affleck and Ludwig \affleckludwig. The connection
between D-brane energy and boundary entropy was made in \callkleb\
for D-branes moving in flat space. Furthermore, we discovered after
completing this work that our computation of the boundary entropy
for torus compactifications in sec 2.2 below  appeared earlier in \ers.

\newsec{The general argument: Bosonic string}

We consider a spacetime defined by a
closed conformal field theory of the type
$\CC(\IR^{1,25-d}) \otimes \CC_2$.
The first factor is the usual conformal field
theory of $26-d$ free uncompactified bosons and
ghosts. For our purposes it suffices to work in light-cone gauge
in which case we can drop the ghost fields and view the
first factor as the CFT of $24-d$ free bosons.
The second factor is an arbitrary unitary
CFT of $c= \bar c = d$. We will choose $d$ sufficiently large so
that we can view the $Dp$-brane we are interested in as a
D-particle in $26-d$ spacetime dimensions.

\subsec{D-brane Tension}

We consider D-brane boundary states of the form:
\eqn\dbfrom{
\vert x \rara \otimes \vert \alpha \rara
}
Here $ \vert \alpha \rara$ is a boundary state for
$\CC_2$ which is assumed to be part of a Cardy set.
It corresponds to an open string channel
statespace $\CH_{\alpha\alpha}$ as in \zcard.
The first factor  $\vert x \rara$ is a standard
position eigenstate D-brane state constructed
from coherent states of lightcone gauge oscillators.
\eqn\xnorm{
\vert x \rara := \CN_{s.t.}
\int d^{24-d} k e^{i k x} e^{-\sum {1\over n}
\alpha_{-n} \tilde \alpha_{-n} } \vert k \rangle
}
Here $\vert k\ra$ denotes a momentum eigenstate in the closed string
Hilbert space;
the normalization $\CN_{s.t.}$ can be
gleaned from \divecchia\ and involves powers
of $2,\pi, \ell_s$.

The D-brane state \dbfrom\
describes a particle in the $(26-d)$-dimensional
spacetime theory. The formula
for the mass of this particle
 in terms of $\CH_{\alpha\alpha}$ is:
\eqn\bosres{
 ( \ell_s M)^2  =  {1\over (64 \pi)^2}
  {(2\pi \ell_s)^{24-d}\over    G_{26-d} }
  \dim \CH_{\alpha\alpha}
}
where $G_{26-d}$ is the Newton constant in 26-d dimensions,
$\ell_s$ is the string length ($\alpha' = \ell_s^2$)
and $\dim \CH_{\alpha\alpha}$ is the regularized dimension,
defined by
\eqn\defrdim{
\dim \CH_{\alpha\alpha}=
\lim_{\tau \rightarrow 0 } e^{2\pi i c/24 (-1/\tau)}
{\Tr}_{\CH_{\alpha\alpha}} q^{L_0 -c/24}
}
where $q=e^{2 \pi i \tau}$.

In order to prove \bosres\ we will assume that $\CC_2$ is described
by rational conformal field theory (RCFT). We believe that this is
only a technical assumption and that \bosres\ holds more generally.
Some evidence for this will be given later where \bosres\ will be
seen to hold in quasirational theories.

In RCFT we have  isomorphic left and right-moving chiral algebras
$\CA_L=\CA_R=\CA$ which contain the Virasoro algebra and which may in
general be subalgebras of a larger chiral algebra. We denote the
moments of the chiral fields generating $\CA$ by $W_n$. By definition, the
Hilbert space can be decomposed into a finite set of irreps $V_j$ of
$\CA$
\eqn\hrdecom{\CH^{\rm closed} = \oplus_{(j, \tilde j)
 \in {\rm Spec}} V_j \otimes
\tilde V_{\tilde j}}
with Spec labelling the irreps in the spectrum, possibly with
multiplicities $N_{j,\tilde j}$. The characters
\eqn\achar{\chi_i(\tau) = {\rm Tr}_{V_i} q^{L_0 -c/24}}
transform under modular transformations $\tau \rightarrow -1/\tau$
according to
\eqn\sdef{\chi_i(-1/\tau) = \sum_j S_{i}^{~j} \chi_j (\tau)}
For each irrep $V_i$ there is a primary field $\phi_i$ obeying the
fusion algebra
\eqn\fuse{\phi_i \times \phi_j = {N_{ij}}^k \phi_k}
with structure constants related to the S matrix by the
Verlinde formula
\eqn\verl{ {N_{ijk}}  = \sum_\ell {S_{i}^{~\ell} S_{j}^{~\ell} S_{k}^{~\ell}
\over S_{1}^{~\ell}  }. }

Demanding that $\CA$ act in the RCFT with boundary requires that
\eqn\weqx{ (W_n - (-1)^{h_W} \Omega(\bar W_{-n})) | \alpha \rara =0}
where $h_W$ is the conformal dimension (spin) of $W$ and $\Omega$
is an automorphism of $\CA$. Character states $| j \rara$ solving
\weqx\ can be constructed by a slight variant of the argument given
earlier \refs{\ishibashi,\behrend}.
The character states $|j \rara$ of $\CA$ do not in general form
a Cardy set. Since character states form a basis, we can write
possible elements of a Cardy set as
\eqn\pdecomp{\vert \alpha \ra\ra = \sum_j {\psi^{j}_{\alpha}
\over \sqrt{  S_{1}^{~j}  }}\vert j \ra\ra}
where the factor in the denominator has been put in for later
convenience. It is positive, see below.

Using
\eqn\normchr{\la\la j \vert  q_c^{{1\over 2}(L_0 + \tilde L_0 - {c\over
12})} \vert k \ra\ra ~=~\delta_{jk} \chi_{j}( q_c)}
we then have
\eqn\zclos{Z_{\beta \alpha} = \sum_j {\psi_\alpha^j (\psi_\beta^j)^* \over
S_{1}^{~j} } \chi_j(q_c) = \sum_{j,k} {\psi_\alpha^j (\psi_\beta^j)^* \over
S_{1}^{~j}} S_{j}^{~k}  \chi_k(q_o)}

On the other hand, since $\CA$ acts on the open string Hilbert space,
we can also decompose $\CH_{\alpha \beta}$ into $\CA$ irreps so that
\eqn\cdhy{{\rm Tr}_{H_{\alpha \beta}} q_o^{L_0 -c/24} =
 \sum_i (n^i)_{\alpha \beta} \chi_i(q_o)}
with $n^i_{\alpha \beta}$ non-negative integers. Equating \cdhy\ and
\zclos\ then gives Cardy's condition in the context of boundary RCFT:
\eqn\rcard{(n^i)_{\alpha\beta} ~=~\sum_j {S_{i}^{~j} \over S_{1}^{~j}}
\psi^{j}_{\alpha} (\psi^{j}_{\beta})^*}

The solution originally given by Cardy has $\alpha$ running over the
irreps of $\CA$ and $(n^i)_{\alpha \beta}=N^i_{\alpha \beta}$,
$\psi_{\alpha j} = S_{\alpha j}$, which clearly solves \rcard\
using \verl.  However, in general other
solutions will exist \refs{\sagnotti,\fs, \behrend}
(this happens for as simple a system as the
rational circle) and we will only assume
\rcard\ in what follows.

We are now ready to prove \bosres.
The key observation is that
the mass is measured by the
one-point function with the graviton.
The graviton
vertex operator $\epsilon_{\mu\nu} \p X^\mu
\bar \p X^\nu e^{i k \cdot X}  \otimes 1$
is the unit operator in the internal theory $\CC_2$.
Thus we need only
know the coupling of $\vert \alpha \rara$ to the
character of the representation with the
unit operator in order to compute the mass of the D-brane.

The unit operator is in a unique character state
$\vert 1 \rara$ and uniqueness of the vacuum
implies $\langle 0 \vert 1 \rara = 1$. Therefore,
the dependence of the tension on the internal
conformal field theory is exactly
$\psi_\alpha^1 / \sqrt{S_{1 1}} $.
On the other hand, the regularized dimension of $\CH_{\alpha \alpha}$ is
\eqn\regdim{\eqalign{
\dim \CH_{\alpha\alpha} &  = \lim_{q_o \rightarrow 1} q_c^{c/24}
{\rm Tr}_{\CH_{\alpha \alpha}} q_o^{L_0-c/24} \cr
 & = \lim_{q_o \rightarrow 1} q_c^{c/24} \sum_{j,k} 
{S_{k}^{~j} \over S_{1}^{~k}}
|\psi_\alpha^k|^2 \chi_j(q_o) \cr
& = {|\psi_{\alpha}^1|^2 \over S_{11}} \cr}}
where we have used  \cdhy,\rcard, and the fact that $\chi_k(q_c)$
goes like $q_c^{h_k -c/24}$ as $q_c \rightarrow 0$ so that the dominant
contribution comes from the identity representation with $h_1=0$ in
the limit.  This then proves \bosres\ up to overall factors which
can be determined by working on the torus (see below) and comparing
to \jp.

It is easy to show that the dimension
(and hence the tension) is nonzero, at least in
RCFTs. In the
Cardy condition we impose positivity
$(n^j)_{\alpha\beta}\geq 0$ and the vacuum must
appear in $(n^0)_{\alpha \alpha}\geq 1$. Moreover,
the matrix element $S_{j1} \geq 0$. This
latter fact  is  easily proved since the regularized dimension
of the representation $j$ of the chiral
algebra is given by the modular matrix
\dv:
\eqn\modmtx{
\dim \CH_j = \lim_{t \rightarrow 0} {{\Tr}_{\CH_j} e^{- 2\pi t H} \over
{\Tr}_{\CH_1} e^{- 2\pi t H} } =
{S_{j1} \over  S_{11}}
}
This is a limit of positive quantities and hence
nonnegative.\foot{We
are assuming unitarity of the internal CFT ${\CC_2}$ here, which
is reasonable for string theory applications.}
In fact, in RCFT the $S_{j1}$ cannot vanish  because that
leads to inconsistencies in the modular representation.
Indeed the interpretation of this quantity
as a positive
dimension is crucial to the general picture of
RCFT as a generalization of group
theory \refs{\naturality,\ms}.

\noindent{\bf Remarks:}

\item{1.} From the closed string point of view, the limit used to define
the regularized dimension is of course the same as the limit originally
used by Polchinski to compute the tension for D-branes in flat space
\joed. 

\item{2.}  The derivation is valid to leading order in
string perturbation theory, which makes it an $\it exact$ statement
for BPS branes. It would be interesting
to see if there is a sense in which it is true beyond
leading order.  Because the result
seems very natural to us, we conjecture that it
will continue to hold even for nonrational backgrounds.

\item{3.} Note that  \bosres\ behaves nicely
upon inclusion of Chan-Paton spaces
\eqn\cpadd{
\sqrt{\dim \bigl(\CH_{\alpha\alpha} \otimes Mat_N(\IR)\bigr) } =
N \sqrt{\dim (\CH_{\alpha\alpha}) }
}
Actually, in string theory we impose a reality
condition on Chan-Paton factors in $Mat_N(\IC)$
so that Chan-Paton factors take values
in the Hermitian matrices, $H_N$,
but the real dimension is again $\dim H_N = N^2$.

\item{4.} As mentioned earlier, the quantity
${\psi_\alpha^{ ~ 1 } \over  \sqrt{S_{1 1} }}$
is what Affleck and Ludwig call the ``nonintegral
groundstate degeneracy'' in boundary CFT. It has been conjectured,
and established in conformal perturbation theory, that this quantity
descreases along renormalization group flows.
See \refs{\affleckludwig,\affludb,\behrend} for further discussion.

\subsec{Example: Compactification on Tori}

A simple example of the above rule is provided
by the Gaussian model on  $T^d$ with constant background
metric $G_{\mu \nu}$ and two-form $B_{\mu \nu}$. We set $E=G+B$.
(If we pick rational values for the Narain moduli, this
case fits simply into the framework discussed above; in the
general quasirational case we will take some shortcuts below
in showing that the squared tension is given by a regularized
dimension, given boundary states satisfying the Cardy condition.)
The closed conformal field theory is
characterized by a Narain lattice $\Gamma(E)\subset \IR^{d,d}$.
The isomorphism  of $u(1)^d$ left and right
chiral algebras is given by
 $\alpha_n = \CR \cdot \tilde \alpha_{-n}$,
where $\CR\in O(d;\IR)$ is some rotation
matrix  \refs{\callan,\fs,\RS}.

Let
\eqn\lambdadf{
\Lambda = \Gamma(E) \cap \{(p_L;p_R): p_L = \CR \cdot p_R \}
}
We denote the rank of $\Lambda$ by $r$ and the metric tensor
on $\Lambda$ by $ \CG_{ij}$.
By this we mean the metric appearing in the (Euclidean) inner product
$p_L^2+p_R^2=n^T{\CG}n$ for a $d$ dimensional vector
of integers $n$.
Note that $\Lambda$ depends on both $E$ and $\CR$.

Let $\theta$ be a character of $\Lambda$.
Cardy states will be of the form
\eqn\cardy{
\vert \theta; E,\CR \rara = N_{\theta} \sum_{\lambda\in \Lambda}
e^{ 2\pi i \theta \cdot \lambda} e^{-S(\CR)} \vert \lambda \rangle
}
where $S(\CR) = \sum_n \alpha_{-n} \CR \tilde \alpha_{-n}/n$
and
we use the discrete measure for the
momentum eigenvectors.
Let us ask that the single state
$\vert \theta; E,\CR \rara$ form a Cardy state.
We compute
\eqn\diagpi{
\lala \theta \vert
 q_c^{{1 \over 2}(L_0 + \tilde L_0 -c/12)}
\vert \theta \rara =
 {|N_\theta |^2 \over  \eta(q_c)^d}
\sum_{\lambda\in \Lambda} q_c^{\half \lambda_L^2} =
{|N_\theta |^2
\over  \eta(q_c)^d} \sum_{n\in \IZ^r}
e^{-\pi t_c  n^i {\CG}_{ij} n^j}
}
{}From Poisson resummation we get
\eqn\diagpitwo{
\lala \theta \vert
q_c^{{1 \over 2}(L_0 + \tilde L_0 -c/12)}
\vert \theta \rara = t_c^{d/2-r/2} {\vert N_\theta \vert^2
\over \sqrt{\det {\CG}_{ij}}}  {1 \over  \eta(q_o)^d}
\sum_{\hat n\in \IZ^r}
e^{-\pi t_o \hat n^i ({\CG}^{-1})^{ij} \hat n^j}
}
and we thus conclude that   $r = {\rm rank}(\Lambda) = d$.
We interpret this to mean that
 D-branes wrapped on foliating subtori
of $T^d$ cannot form Cardy states.
Moreover, the minimal normalization is $N_\theta =
 (\det {\CG}_{ij})^{1/4}$ and
therefore, the overlap with the unit $u(1)^d$ character
is $N_\theta =  (\det {\CG}_{ij})^{1/4}$. Equivalently,
picking the tension out of the leading piece as
$q_c\to 0$ in the closed string channel,
\eqn\dimspc{
\dim\CH_{\theta\theta} = (\det {\CG}_{ij})^{1/2}
}

For example, choosing the diagonal torus
with Narain lattice
$\Gamma(E) = \{  {1 \over  \sqrt{2} }({n_i \over  R_i} - m_i R_i;
{n_i \over  R_i} +  m_i R_i) \}
$
and rotation
$\CR = diag\{ -1^p; +1^{d-p}\} $ corresponding to a
wrapped $p$-brane then one easily finds:
\eqn\dimspc{
\dim\CH_{\theta\theta} =\prod_{i=1}^p R_i  \prod_{i=p+1}^d
{1\over  R_i}
}

If one adds a flat $B$-field and considers a $d$-brane
wrapping $T^d$ then
 $\CR = -  E^{-1} E^{tr} $. In this
case one finds
\eqn\dimbee{
\dim\CH_{\theta\theta} = (\det {\CG}_{ij})^{1/2} =
\vert \det (G_{\mu\nu} + B_{\mu\nu}) \vert ( \det G_{\mu \nu})^{-1/2}.
}

Both of these agree with standard formulae for the mass of a wrapped
brane when one recalls that $1/G_{26-d}= \sqrt{\det G_{\mu \nu}}/G_{26}$.
In particular, \dimbee\
gives the sensible result that the tension is
given by the Born-Infeld action
$\sqrt{\vert \det (G_{\mu\nu} + B_{\mu\nu} ) \vert}$,
a result which is essentially already to be
found in \callan.

\newsec{The general argument: Superstrings}

\subsec{Cardy condition for $\CN=1$ superconformal
field theory}

To define the superconformal algebra we need
to pick a spin structure. The index set $\CI$ of
characters divides into NS and R sectors which
we denote as:
$\CI_+:=\CI_{NS}  $, $\CI_-:= \CI_{R}   $.
 Similarly,
we must consider both characters and characters
twisted by $(-1)^F$:
\eqn\twischrs{
\chi_i^\epsilon(q) := {\Tr}_{\CH_i} (\epsilon^F q^H)
}
where $\epsilon=\pm 1$.
The transformation $\tau \rightarrow -1/\tau$
acts in a standard way on the spin structures,
so we can define several modular matrices:
\eqn\modmatrs{
\eqalign{
\chi_i^{+}(q_c) & = \sum_{j\in \CI_+} S^{++}_{ij} \chi_j^{+}(q_o) \qquad
i\in\CI_+\cr
\chi_i^{-}(q_c) & = \sum_{j\in \CI_-} S^{+-}_{ij} \chi_j^{+}(q_o) \qquad
i\in\CI_+\cr
\chi_i^{+}(q_c) & = \sum_{j\in \CI_+}
S^{-+}_{ij} \chi_j^{-}(q_o) \qquad i\in\CI_-\cr
\chi_i^{-}(q_c) & = \sum_{j\in \CI_-}
S^{--}_{ij} \chi_j^{-}(q_o) \qquad i\in\CI_-\cr}
}
Note that $(S^{++})^2=C$ is the conjugation matrix
(and for simplicity we will now assume reps are
self-conjugate).
$S^{- -}_{ij}=\delta_{ij}$ because
the Witten index is
modular invariant. \foot{
 Warning: We are using a possibly confusing
piece of notation. $\chi_i^+$ for $i\in \CI_+$ is the
path integral for the spin structure commonly
denoted $(-,-)$!}

We can form character states $\vert i; \eta \rara$
in the standard way. $\eta$ denotes the choice
of isomorphism between left and right supercurrents
so the character basis gives a basis of solutions to
the linear equations
\eqn\dfneta{
(G_r - i \eta \tilde G_{-r} ) \vert\alpha \rara=0
}
where $r \in \IZ$ in the R  sector and $r \in \IZ+1/2$ in the
NS sector.

We can form Cardy states in the usual way:
\eqn\crdysts{
\vert  \alpha, \epsilon ; \eta \rara
= \sum_{i\in \CI_\epsilon} {\psi_\alpha^{~i}(\eta) \over
\sqrt{S_{i1}^{\epsilon,+} }  }  \vert i; \eta \rara
}

The Cardy condition becomes
\eqn\supermumbo{
\lala \alpha, \epsilon; \eta \vert q_c^{{1 \over 2}(L_0 + \tilde L_0
-{c \over 12})}
 \vert \beta, \epsilon; \eta' \rara
={\Tr}_{\CH_{\alpha\beta}^{\eta\eta' } }
\epsilon^F q_o^{H_o}= \sum_{j\in \CI_{\eta\eta' } } (n_j^{\epsilon, \eta\eta'}
)_\beta^{~~\alpha} \chi_j^\epsilon(q_o)
}
In a standard unitary theory we will require  that
$(n_j^{\epsilon,\eta\eta'})_\alpha^{~\beta}\geq 0$.

In particular, focusing on the $(-,-)$ spin structure,
which is invariant under $\tau \rightarrow -1/\tau$,  we have:
\eqn\superjumbo{
(n_i^{++})_\alpha^{~ \beta} = \sum_{j\in \CI_+ }
 {S_{ij}^{++} \over  S_{1j}^{++} }
\psi_\alpha^j(\eta)  (\psi_\beta^j(\eta)  )^*
}
We find again that
\eqn\regdimsup{
\dim \CH_{\alpha\alpha}^{NS}=
\dim \CH_{\alpha,\alpha }^{\eta,\eta} = \left({\psi_\alpha^{~1 }(\eta) \over
\sqrt{S_{11}^{++} } }\right)^2 = \sum_{j\in \CI_+} (n_j^{++})_\alpha^{~~\alpha}
S^{++}_{j1}
}

\noindent{\bf Remarks:}

\item{1. }  Equation \dfneta\ is only one
of several   choices one must make in
the  isomorphism of left and right chiral algebras.

\item{2.} In \crdysts\ we
are considering a Cardy set with states with fixed
$\eta$ and
which are either purely NS or purely R. To make GSO invariant
states it is necessary to combine states with different
$\eta$ as described below.  We can
also form Cardy sets with  states which are
linear combinations of character states from
the R and NS sector, and this is required in order to obtain
BPS boundary states. There are only minor modifications
to the analysis.

\item{3.} In \regdimsup\
 $\psi_\alpha^{~1 }(\eta)$ is independent of $\eta$.

\item{4.}  In \regdimsup\
 $n_j^{++}\geq 0$, and if the unit operator is
present then $n_j^{++}$ is positive definite for $j=0$.
 Thus, as for the bosonic
string, we can conclude that the right hand side is positive.

\subsec{Tension and dimension for the superstring}

We begin with a product of superconformal
theories $\CC_1(\IR^{8-d}) \otimes \CC_2$
where $\CC_1(\IR^{8-d})$ is the lightcone
superconformal field theory of $8-d$
free bosons and fermions $X^\mu, \psi^\mu$.
$\CC_2$ is a unitary superconformal
field theory with $\hat c = \hat { \bar c} = d $.
The closed string is a subtheory because
(a) we restrict to the subspace with the
same spin structures on both factors and
(b) we GSO project. The statespace has
the form:
\eqn\statespace{
(\CH^1_{NS} \otimes \CH^2_{NS})^+
\oplus
(\CH^1_{R} \otimes \CH^2_{R})^+ .
}
where the $+$ superscript indicates the need to take a GSO projection.
We will take the brane to be a point particle in
the uncompactified dimensions. The general
form of the boundary state is then \bg\
\eqn\genform{
\eqalign{
C^{\alpha, NS,+}
\vert x, NS; \eta \rara^{(1)} \otimes \vert \alpha, NS; \eta \rara^{(2)}
 & + C^{\alpha,NS,-}
\vert x, NS; -\eta \rara^{(1)} \otimes \vert \alpha, NS; -\eta \rara^{(2)}
 \cr
+
C^{\alpha, R,+}
\vert x, R; \eta \rara^{(1)} \otimes \vert \alpha, R; \eta \rara^{(2)}
& + C^{\alpha,R,-}
\vert x, R; -\eta \rara^{(1)} \otimes \vert \alpha, R; -\eta \rara^{(2)}
 \cr}
}

Here we will assume the state is made from a Cardy set
of boundary  states of the internal theory, although
this is probably not necessary (i.e., only the combined
theory really needs to obey sewing axioms).
The constants are determined by imposing the GSO projection, by
choosing BPS or anti-BPS branes,and
by Cardy's condition. In supersymmetric type II theory the GSO
projection requires
 $ C^{\alpha,NS,-}
=-C^{\alpha, NS,+} $ and $C^{\alpha,R,+}=C^{\alpha,R,-}$. Note that
the GSO projection does
{\it not} project out the unit operator in the
internal theory. Other choices of GSO projection will lead to other
conditions  on the coefficients. For example, in type 0 theory where
one has a diagonal GSO projection, $C^{\alpha, NS,+}$
and $C^{\alpha,NS,-}$ are independent, reflecting the doubling of the
number of D-brane states \bcr.

The graviton vertex operator is $\epsilon_{\mu\nu} \psi^\mu \tilde\psi^\nu
e^{ik X} \otimes 1$ so the coupling of the
state \genform\ to the graviton is, up to a phase,
\eqn\coupling{
  (C^{\alpha, NS,+ } -C^{\alpha, NS,-})
{\psi_\alpha^{~1 }(\eta) \over
\sqrt{S_{11}^{++} } }
}

We can obtain a quantization condition on the
coefficient $C^{NS}$ in
order to ensure that the combined state continues
to satisfy the Cardy condition. The condition
depends on whether or not we want to enforce the
GSO projection in the open string channel (i.e.
whether or not we consider a BPS or a non-BPS
type brane). If we do not enforce the
condition then $C^R=0$ and
$2 \vert C^{NS} \vert^2 \in \IZ_+$.
Taking the minimal value allowed we get
the open string NS sector is
\eqn\osns{
{\Tr}_{\CH_{\alpha\alpha}^{NS} } q_o^{H_o} = { 1\over  \eta^{(8-d)/2} }
\biggl({\vartheta_3 \over  \eta}\biggr)^{(8-d)/2} \sum_{i\in \CI_+}
n^i_{\alpha\alpha} \chi_i^+(q_o)
}
If we want to have the usual  GSO projection then
we take a bilinear form for the cylinder amplitude
\foot{Thus changing the orientation so that
we have two ingoing circles. This is necesary
to get the spacetime fermion minus sign. } and
impose   $2 (C_{12}^{NS})^2 = -  (C_{12}^R)^2/8 = {n\over  2}$ and taking the
minimal value $n=1$ we have
\eqn\osnsp{
{\Tr}_{\CH_{\alpha\alpha}^{NS} } q_o^{H_o} =
{ 1\over  \eta^{(8-d)/2} }
\half \Biggl[
\biggl({\vartheta_3 \over  \eta}\biggr)^{(8-d)/2}
\sum_{i\in \CI_+} n^i_{\alpha\alpha} \chi_i^+(q_o)
+ \biggl({\vartheta_4 \over  \eta}\biggr)^{(8-d)/2}  \sum_{i\in \CI_+}
n^i_{\alpha\alpha} \chi_i^-(q_o) \Biggr]
}

In either case, taking into account \coupling\
and fixing the overall normalization from the
torus case we get the superanalog of \bosres:
\eqn\superes{
(\ell_s M)^2 = {1 \over  16\pi^2}  {(2\pi \ell_s)^{8-d} \over  G_{10-d} }
 \dim \CH_{\alpha\alpha}^{NS}
}
Here  $\CH_{\alpha\alpha}^{NS} $ is the open string
channel NS sector of the full string theory.
In particular, this
  result holds for BPS and nonBPS,
the difference of a factor of $\sqrt{2}$ in the
tension between a BPS and non-BPS brane \sennon\
is due to the factor of $1/2$ in the projection
operator which reduces  the dimension of $\CH_{\alpha \alpha}^{NS}$
by a factor of two.

\newsec{Discussion}

We conclude with two speculative remarks
on possible future applications of this work.

First, it is natural to speculate that the mass formula derived here
in terms of the Affleck-Ludwig degeneracy $g$
\eqn\massfo{\CM^2 \sim \vert g \vert^2
 = {\Tr}_{\CH_{\alpha\alpha}}(1)}
could play a role in extensions of the attractor mechanism \fks\ to
nonsupersymmetric, spherically symmetric charged black holes.
In the context of BPS black holes in $\CN=2$ supersymmetric
compactifications to
four dimensions, it is clear that $g$ can be identified with the
central charge $\vert \CZ\vert$.  The attractor mechanism tells us that we
should associate spherically symmetric
extremal RR charged black holes with gradient flow on the
moduli space using the Zamolodchikov
metric and the potential
${\rm log  \vert \CZ\vert} \sim {\rm log(g)}$
\refs{\fgk,\arthatt}.
Since $g$ is intrinsically well defined in string theory without
any reference to supersymmetry, one can conjecture that
in general compactifications,
one can associate spherically symmetric extremal RR charged black hole
solutions with gradient flows using the
Zamolodchikov metric and the potential
${\rm log(g)}$.  The ``doubly extremal'' solutions (with constant values of
the closed string moduli as a function of the radius) then arise
when one chooses
the moduli to sit at a (local) minimum of $g$. It would
also be interesting to see if the dynamical evolution
of couplings on a
test 3-brane falling into a black hole is related to
renormalization group flow of the boundary entropy.

Second, our result fits in well with the currently
emerging interconnections between
D-branes, K-theory, and noncommutative
geometry.
In the framework of noncommutative
geometry formulae have been derived
for the energy of wrapped D-branes in
\cds\ and subsequent papers. See
\piosch\ for a recent discussion. In
order to account for the mass $\CM$
of a wrapped D-brane in its groundstate
one must add to the noncommutative
Yang-Mills action a term proportional
to ${\Tr}_E(1)$, where $E$ is the projective
module of sections of the (noncommutative)
Chan-Paton bundle. On the other hand,
in  the present paper
we have derived a relation of the
form $\CM^2 \sim {\Tr}_{\CH_{\alpha\alpha}}(1)$.
The two results are compatible if we
can identify (at least in the $\alpha^\prime\to 0$ limit
corresponding to the NCSYM theory)  $\CH_{\alpha\alpha} = {\rm End}(E)$,
and such an identification in this limit is strongly
suggested by the behavior of the finite dimensional
Chan-Paton factors. Clearly, this connection
deserves closer scrutiny.

It is also worth noting that in the theory of
operator algebras one can define the
Murray-von Neumann dimension,\foot{
In the literature on operator algebras this
is also  called the ``coupling constant.''}
which assigns a continuous dimension to
certain Hilbert spaces of operators. It is
well-known that, at least in some RCFTs,
the regularized dimensions \modmtx\ are
indeed given by ratios of such
Murray-von Neumann dimensions.
In this way the Jones index of subfactors
appears in RCFT. On the other hand, these
same dimensions can be related to traces of
projection operators on  towers
of finite-dimensional multimatrix algebras
\ghj. Analogous constructions applied to
the ``algebra'' of open string vertex operators
might lead to interesting new ways of formulating
D-branes, and, perhaps, even $M$-theory.

\bigskip
\centerline{\bf Acknowledgements}\nobreak
\bigskip

We would like to thank
T. Banks, J. Cardy, M. Douglas, P. Fendley, D. Friedan,  J. Maldacena,
A. Recknagel,
H. Saleur, V. Schomerus, N. Seiberg, I. Singer and E. Witten
for useful discussions and remarks.
We would like to acknowledge the hospitality of
the Aspen Center for Physics and the
Amsterdam Summer Workshop on String Theory.
GM would like to thank the Institute
for Advanced Study for hospitality and
the Monell foundation for support.
The  work  of JH is supported by
NSF Grant No.~PHY 9901194, SK is supported by an
A.P. Sloan Foundation Fellowship and a DOE OJI Award, GM is
supported by
DOE grant DE-FG02-92ER40704, and ES is supported by an A.P. Sloan
Foundation Fellowship, a DOE OJI Award, and
by the DOE under contract DE-AC03-76SF00515.

\listrefs

\bye